\documentclass[journal]{IEEEtran}
\usepackage{cite}
\usepackage{graphicx}
\usepackage{amssymb,amsmath,bm}
\usepackage[colorlinks,linkcolor=black,anchorcolor=black,citecolor=black,urlcolor=black]{hyperref}
\usepackage{hyperref}
\usepackage{multirow}
\usepackage{tabularx}
\newcommand{\tabincell}[2]{\begin{tabular}{@{}#1@{}}#2\end{tabular}}

\ifCLASSINFOpdf
\else
\fi
\begin{document}
%
\title{Waveform Modeling and Generation Using Hierarchical Recurrent Neural Networks for Speech Bandwidth Extension}
%
%
%

\author{Zhen-Hua~Ling,~\IEEEmembership{Member,~IEEE},~Yang~Ai,~Yu~Gu,~and~Li-Rong~Dai
\thanks{This work was partially funded by National Key Research and Development Project of China (Grant No. 2017YFB1002202)
and the National Natural Science Foundation of China (Grants No. U1636201).}
\thanks{Z.-H. Ling, Y. Ai, and L.-R. Dai are with the National Engineering Laboratory of Speech and Language Information Processing,
University of Science and Technology of China, Hefei, 230027, China (e-mail: zhling@ustc.edu.cn, ay8067@mail.ustc.edu.cn, lrdai@ustc.edu.cn).}
\thanks{Y. Gu is with Baidu Speech Department, Baidu Technology Park, Beijing, 100193, China (e-mail: guyu04@baidu.com ). This work was done when he was
a graduate student at the National Engineering Laboratory of Speech and Language Information Processing,
University of Science and Technology of China. }
}

%
%

\markboth{Journal of \LaTeX\ Class Files,~Vol.~6, No.~1, January~2007}%
{Shell \MakeLowercase{\textit{et al.}}: Bare Demo of IEEEtran.cls for Journals}
%



\maketitle

\begin{abstract}
This paper presents a waveform modeling and generation method using hierarchical recurrent neural networks (HRNN) for speech bandwidth extension (BWE).
Different from conventional BWE methods which predict spectral parameters 
for reconstructing wideband speech waveforms,
this BWE method models and predicts waveform samples directly without using vocoders.
Inspired by SampleRNN which is an unconditional neural audio generator,
the HRNN model represents the distribution of each wideband or high-frequency waveform sample conditioned on the input narrowband waveform samples
using a neural network composed of long short-term memory (LSTM) layers and feed-forward (FF) layers.
The LSTM layers form a hierarchical structure and each layer operates at a specific temporal resolution to efficiently capture long-span dependencies between temporal sequences.
Furthermore, additional conditions, such as the bottleneck (BN) features derived from narrowband speech using a deep neural network (DNN)-based state classifier, are employed as auxiliary input to further improve the quality of generated wideband speech.
The experimental results of comparing several waveform modeling methods show that the HRNN-based method can achieve better speech quality and run-time efficiency than the dilated convolutional neural network (DCNN)-based method and
the plain sample-level recurrent neural network (SRNN)-based method.
Our proposed method also outperforms the conventional vocoder-based BWE method using LSTM-RNNs in terms of the subjective quality of the reconstructed wideband speech.

\end{abstract}

\begin{IEEEkeywords}
speech bandwidth extension, recurrent neural networks, dilated convolutional neural networks, bottleneck features
\end{IEEEkeywords}

%
\IEEEpeerreviewmaketitle

\section{Introduction}
\label{sec1: Introduction}
\IEEEPARstart{S}
{peech} communication is important in people's daily life. 
However, due to the limitation of transmission channels and the restriction of speech acquisition equipments,
the bandwidth of speech signal is usually limited to a narrowband of frequencies.
For example, the bandwidth of speech signal in the public switching telephone network (PSTN) is less than 4kHz.
The missing of high-frequency components of speech signal usually leads to low naturalness and intelligibility, such as the difficulty of distinguishing fricatives and similar voices.
Therefore, speech bandwidth extension (BWE), which aims to restore the missing high-frequency components of narrowband speech using the correlations that exist between
the low and high-frequency components of the wideband speech signal,
has attracted the attentions of many researchers.
BWE methods can not only be applied to real-time voice communication, but also benefit other speech signal processing areas such as text-to-speech (TTS) synthesis \cite{nakamura2014mel},  speech recognition \cite{albahri2016artificial,goodarzi2012feature}, and speech enhancement \cite{chennoukh2001speech,mustiere2010bandwidth}.

Many researchers have made a lot of efforts in the field of BWE.
Some early studies adopted the source-filter model of speech production and attempted to restore high-frequency residual signals and spectral envelopes respectively from input narrowband signals.
The high-frequency residual signals were usually estimated from the narrowband residual signals by spectral folding \cite{makhoul1979high}.
To estimate high-frequency spectral envelopes from narrowband signals is always a difficult task. 
To achieve this goal, simple methods, such as codebook mapping \cite{vaseghi2006speech} and linear mapping \cite{chennoukh2001speech}, and statistical methods using Gaussian mixture models (GMMs) \cite{pulakka2011speech,wang2015speech,ohtani2014gmm,zhang2009speech} and hidden Markov models (HMMs)\cite{song2009study,yong2014bandwidth,bauer2008hmm,chen2004hmm}, have been proposed.
In statistical methods, acoustic models were build to represent the mapping relationship between narrowband spectral parameters 
and high-frequency spectral parameters. 
Although these statistical methods achieved better performance than simple mapping methods,
the inadequate modeling ability of GMMs and HMMs may lead to over-smoothed spectral parameters 
which constraints the quality of reconstructed speech signals \cite{ling2015deep}.

In recent years, deep learning has become an emerging field in machine learning research. Deep learning techniques have been successfully applied to many signal processing tasks.
In speech signal processing, neural networks with deep structures have been introduced to the speech generation tasks including speech synthesis \cite{ling2013modeling, zen2013statistical},
voice conversion \cite{chen2014voice,nakashika2013voice}, speech enhancement \cite{lu2013speech,xu2015regression}, and so on.
In the field of BWE, neural networks have also been adopted to predict either the spectral parameters 
representing vocal-tract filter properties \cite{botinhao2006frequency,kontio2007neural,pulakka2011bandwidth}
or the original log-magnitude spectra derived by short-time Fourier transform (STFT)  \cite{li2015deep,liu2015novel}.
The studied model architectures included deep neural networks (DNN) \cite{wang2015speech2,abel2016artificial,gu2015restoring}, recurrent temporal restricted Boltzmann machines (RBM) \cite{wang2016speech},
 recurrent neural networks (RNN) with long short-term memory (LSTM) cells \cite{gu2016speech}, and so on.
These methods achieved better BWE performance than using conventional statistical models, like GMMs and HMMs, since deep-structured neural networks are more capable of modeling  the complicated and nonlinear mapping relationship between input and output acoustic parameters. 

However, all these existing methods are vocoder-based ones,
which means vocoders are used to extract spectral parameters 
from narrowband waveforms and then to reconstruct waveforms from the predicted wideband or high-frequency spectral parameters. 
This may lead to two deficiencies.
First, the parameterization process of vocoders usually degrades speech quality.
For example, the spectral details are always lost in the reconstructed waveforms when low-dimensional spectral parameters, such as mel-cepstra or line spectral pairs (LSP), are adopted to represent spectral envelopes in vocoders.
The spectral shapes of the noise components at voiced frames are always ignored when only F0 values and binary voiced/unvoiced flags are used to describe the excitation.
Second, it is difficult to parameterize and to predict phase spectra due to the phase-warpping issue.
Thus, simple estimation methods, such as mirror inversion, are popularly used to predict the high-frequency  phase spectra in existing methods \cite{li2015deep,gu2016speech}.
This also constraints the quality of the reconstructed wideband speech.

Recently, neural network-based speech waveform synthesizers, such as  WaveNet \cite{oord2016wavenet} and SampleRNN \cite{mehri2016samplernn}, have been presented.
In WaveNet \cite{oord2016wavenet}, the distribution of each waveform sample conditioned on previous samples and additional conditions was represented
using a neural network with dilated convolutional neural layers and residual architectures.
SampleRNN \cite{mehri2016samplernn} adopted recurrent neural layers with a hierarchical structure for unconditional audio generation.
Inspired by WaveNet, a waveform modeling and generation method using stacked dilated CNNs for BWE has been proposed in our previous work \cite{gu2017speech},
which achieved better subjective BWE performance than the vocoder-based approach utilizing LSTM-RNNs.
On the other hand, the methods of applying RNNs to directly model and generate speech waveforms for BWE have not yet been investigated.


Therefore, this paper proposes a waveform modeling and generation method using RNNs for BWE.
As discussed above, direct waveform modeling and generation can help avoid the spectral representation and phase modeling issues in vocoder-based BWE methods.
Considering the sequence memory and modeling ability of RNNs and LSTM units,
this paper adopts LSTM-RNNs to model and generate the wideband or high-frequency waveform samples directly given input narrowband waveforms.
Inspired by SampleRNN \cite{mehri2016samplernn},  a hierarchical RNN (HRNN) structure is presented for the BWE task.
There are multiple recurrent layers  in an HRNN and each layer operates at a specific temporal resolution.
Compared with plain sample-level deep RNNs, HRNNs are more capable and efficient at capturing long-span dependencies in temporal sequences.
Furthermore, additional conditions, such as the bottleneck (BN) features \cite{gu2016speech,yu2011improved,wu2015deep} extracted from narrowband speech using a DNN-based state classifier,
are introduced into HRNN modeling to further improve the performance of BWE.

The contributions of this paper are twofold.
First, this paper makes the first successful attempt to model and generate speech waveforms directly at sample-level using RNNs for the BWE task.
Second, various RNN architectures for waveform-based BWE, including plain sample-level LSTM-RNNs, HRNNs, and HRNNs with additional conditions, are implemented and evaluated  in this paper.
The experimental results of comparing several waveform modeling methods show that the HRNN-based method achieves better speech quality and run-time efficiency than the stacked dilated CNN-based method \cite{gu2017speech} and the plain sample-level RNN-based method.
Our proposed method also outperforms the conventional vocoder-based BWE method using LSTM-RNNs in terms of the subjective quality of the reconstructed wideband speech.

This paper is organized as follows. In Section \ref{sec2: Previous Work}, we briefly review previous BWE methods including vocoder-based ones and the dilated CNN-based one.
In Section \ref{sec3: Proposed Method}, the details of our proposed method are presented.
Section \ref{sec4: Experiments} reports our experimental results, and conclusions are given in Section \ref{sec5: Conclusion}.

\section{Previous Work}
\label{sec2: Previous Work}

\subsection{Vocoder-Based BWE Using Neural Networks}
\label{subsec2B: Vocoder-based BWE Using LSTM-Based RNNs}

The vocoder-based BWE methods using DNNs or RNNs have been proposed in recent years \cite{li2015deep,gu2016speech}.
In these methods, spectral parameters 
such as logarithmic magnitude spectra (LMS) were first extracted by short time Fourier transform (STFT) \cite{allen1977unified}.
Then, DNNs or LSTM-RNNs were trained under minimum mean square error (MMSE) criterion to establish a mapping relationship from the LMS of narrowband speech to the LMS of the high-frequency components of wideband speech.
Some additional features extracted from narrowband speech, such as bottleneck features, can be used as auxiliary inputs to improve the performance of networks \cite{gu2016speech}.
At the stage of reconstruction, the LMS of wideband speech were reconstructed by concatenating the LMS of input narrowband speech and the LMS of high-frequency components predicted by the trained DNN or LSTM-RNN.
The phase spectra of wideband speech were usually generated by some simple mapping algorithms, such as mirror inversion \cite{li2015deep}.
Finally, inverse FFT (IFFT) and overlap-add algorithm were carried out to reconstruct the wideband waveforms from the predicted LMS and phase spectra.

The experimental results of previous work showed that LSTM-RNNs can achieve better performance than DNNs in the vocoder-based BWE \cite{gu2016speech}.
Nevertheless, there are still some issues with the vocoder-based BWE approach as discussed in Section \ref{sec1: Introduction}, such as the quality degradation caused by the parameterization of vocoders and the inadequacy of restoring phase spectra.

\subsection{Waveform-Based BWE Using Stacked Dilated CNNs}
\label{subsec2C: Waveform-Based BWE Using Stacked Dilated CNNs}

Recently, a novel waveform generation model named WaveNet  was proposed \cite{oord2016wavenet} and has been successfully applied to the speech synthesis task \cite{arik2017deep,tamamori2017speaker,hu2017ustc}.
This model utilizes stacked dilated CNNs to describe the autoregressive generation process of audio waveforms without using frequency analysis and vocoders.
A stacked dilated CNN consists of many convolutional layers with different dilation factors. The length of its receptive filed grows exponentially in terms of the network depth \cite{oord2016wavenet}.

\begin{figure}[t]
    \centering
    \includegraphics[height=4.5cm]{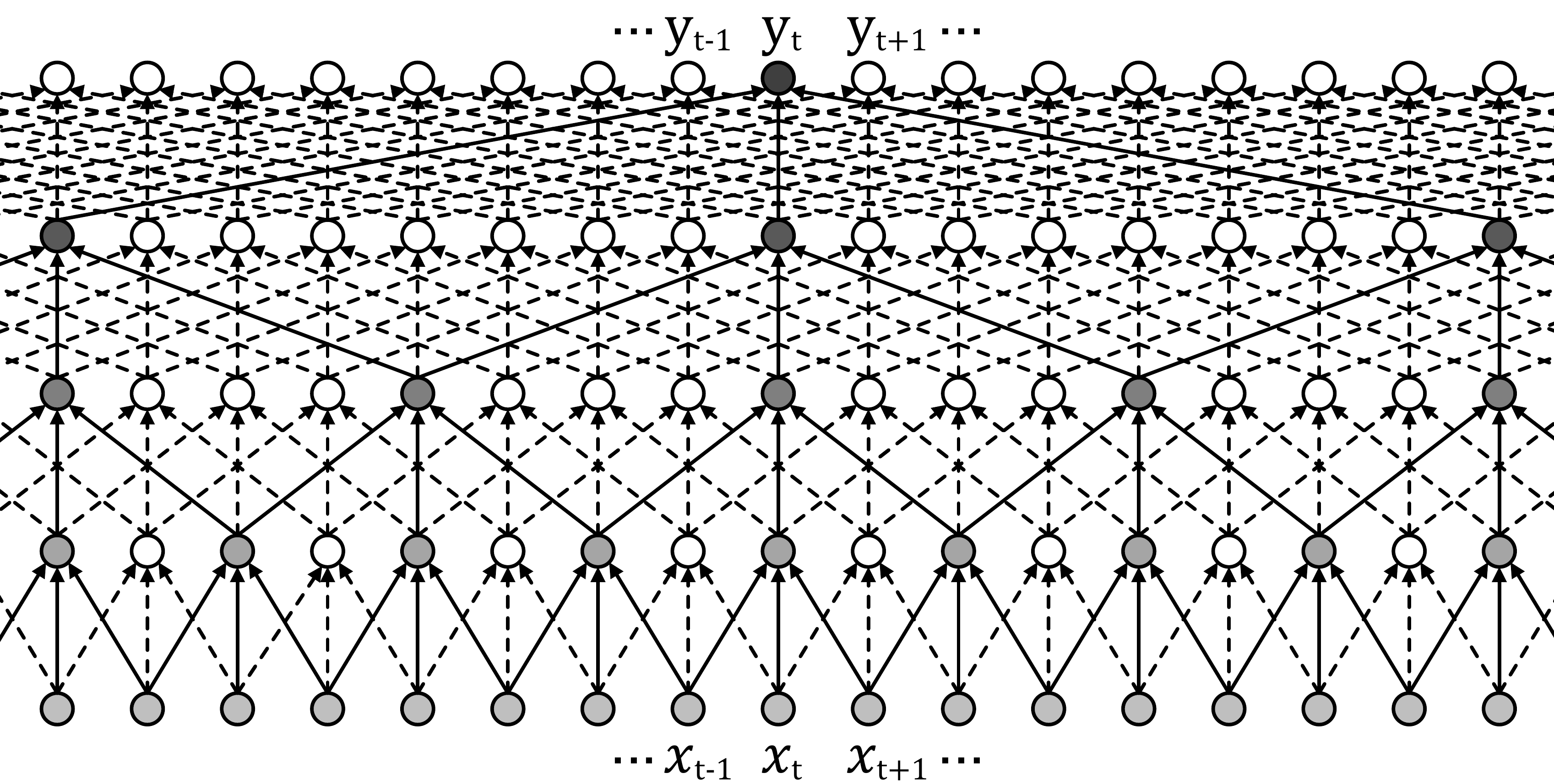}
    \caption{The structure of stacked dilated non-causal CNNs \cite{gu2017speech}.}
    \label{fig2: DCNNs}
\end{figure}

Motivated by this idea, a waveform modeling and generation method for BWE was proposed \cite{gu2017speech},
which described the conditional distribution of the output wideband or high-frequency waveform sequence $\bm{y}=[y_1,y_2,\dots,y_T]$ conditioned on the input narrowband waveform sequence $\bm{x}=[x_1,x_2,\dots,x_T]$
using stacked dilated CNNs .
Similar to WaveNet, the samples $x_t$ and $y_t$ were all discretized by 8-bit $\mu$-law quantization \cite{recommendation1988g} and a softmax output layer was adopted.
Residual and parameterized skip connections together with gated activation functions were also employed to capacitate training deep networks and to accelerate the convergence of model estimation.
Different from WaveNet, this method modeled the mapping relationship between two waveform sequences, not the autoregressive generation process of output waveform sequence.
Both causal and non-causal model structures were implemented and experimental results showed that the non-causal structure achieved better performance than the causal one \cite{gu2017speech}.
The stacked dilated non-causal CNN, as illustrated in Fig. \ref{fig2: DCNNs},  described the conditional distribution as
\begin{align}
\label{equ: conditional distribution of DCNNs}
p(\bm{y}|\bm{x})=\prod_{t=1}^Tp(y_t|x_{t-N/2},x_{t-N/2+1},\dots,x_{t+N/2}),
\end{align}
where $N+1$ is the length of receptive field.

At the extension stage, given input narrowband speech, each output sample was obtained by selecting the quantization level with maximum posterior probability.
Finally, the generated waveforms were processed by a high-pass filter and then added with the input narrowband waveforms to reconstruct the final wideband waveforms.
Experimental results showed that this method achieved better subjective BWE performance than the vocoder-based method using LSTM-RNNs \cite{gu2017speech}.

\section{Proposed Methods}
\label{sec3: Proposed Method}


Inspired by SampleRNN \cite{mehri2016samplernn} which is an unconditional audio generator containing recurrent neural layers with a hierarchical structure,
this paper proposes waveform modeling and generation methods using RNNs for BWE.
In this section, we first introduce the plain sample-level RNNs (SRNN) for waveform modeling.
Then the structures of hierarchical RNNs (HRNN) and conditional HRNNs are explained in detail.
Finally, the flowchart of BWE using RNNs is introduced.

\subsection{Sample-Level Recurrent Neural Networks}
\label{subsec3A: Sample-Level Recurrent Neural Networks}

\begin{figure}[t]
    \centering
    \includegraphics[height=5cm]{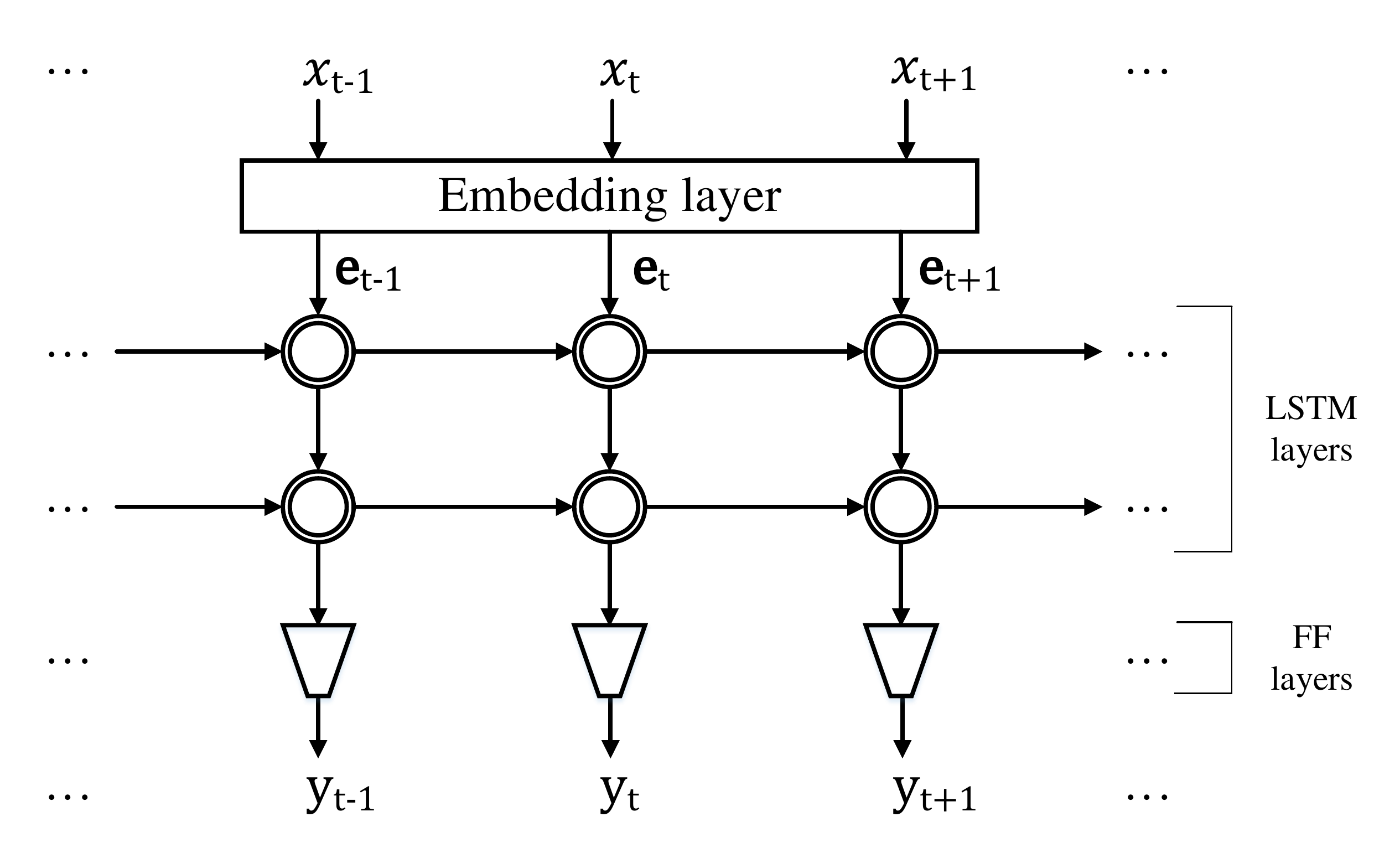}
    \caption{The structure of SRNNs for BWE, where concentric circles represent LSTM layers and inverted trapezoids represent FF layers.}
    \label{fig4: SRNNs}
\end{figure}

The LSTM-RNNs for speech generation are usually built at frame-level in order to model the acoustic parameters extracted by vocoders with a fixed frame shift \cite{fan2014tts,gu2016speech}.
It is straightforward to model and generate speech waveforms at sample-level using similar LSTM-RNN framework.
The structure of sample-level recurrent neural networks (SRNNs) for BWE is shown in Fig. \ref{fig4: SRNNs}, which is composed of a cascade of LSTM layers and feed-forward (FF) layers.
Both the input waveform samples $\bm{x}=[x_1,x_2,\dots,x_T]$ and output waveform samples  $\bm{y}=[y_1,y_2,\dots,y_T]$ are quantized to discrete values by $\mu$-law.
The embedding layer maps each discrete sample value $x_t$  to a real-valued vector $\bm{e}_t$.
The LSTM layers model the sequence of embedding vectors in a recurrent manner. 
When there is only one LSTM layer, the calculation process can be formulated as
\begin{align}
\label{equ: Computation of RNNs}
\bm{h}_t={\cal H}(\bm{h}_{t-1},\bm{e}_t),
\end{align}
where $\bm{h}_t$ is the output of LSTM layers at time step $t$, $\cal H$ represents the activation function of LSTM units.
If there are multiple LSTM layers, their output can be calculated layer-by-layer.
Then, $\bm{h}_t$ passes through FF layers.
The activation function of the last layer is a softmax function which generates the probability distribution of the output sample $y_t$ conditioned on the previous and current input samples $\{x_1,x_2,\dots,x_t\}$ as
\begin{align}
\label{equ: Conditional probability of SRNNs}
p(y_t|x_1,x_2,\dots,x_t)=FF(\bm{h}_t),
\end{align}
where function $FF$ denotes the calculation of FF layers.

Given a training set with parallel input and output waveform sequences, the model parameters of the LSTM and the FF layers are estimated using cross-entropy cost function.
At generation time, each output sample $y_t$ is obtained by maximizing the conditional probability distribution (\ref{equ: Conditional probability of SRNNs}).
Our preliminary and informal listening test showed that this generation criterion can achieve better subjective performance than generating random samples from the distribution.
The random sampling is necessary for the conventional WaveNet and SampleRNN models because of their autoregressive architecture.
However, the model structure shown in Fig. \ref{fig4: SRNNs} is not an autoregressive one.
The input waveforms provide the necessary randomness to synthesize the output speech, especially the unvoiced segments.

In an SRNN, the generation of each output sample depends on all previous and current input samples.
However, this plain LSTM-RNN architecture still has some deficiencies for waveform modeling and generation.
First, sample-level modeling makes it difficult to model long-span dependencies between input and output speech signals due to the significantly increased sequence length compared with frame-level modeling.
Second, SRNNs suffer from the inefficiency of waveform generation due to the point-by-point calculation at all layers  and the dimension expansion at the embedding layer.
Therefore, inspired by SampleRNN \cite{mehri2016samplernn}, a hierarchical RNN (HRNN) structure is proposed in next subsection to alleviate these problems.
\subsection{Hierarchical Recurrent Neural Networks}
\label{subsec3B: Hierarchical Recurrent Neural Networks}

\begin{figure}[t]
    \centering
    \includegraphics[height=5.2cm]{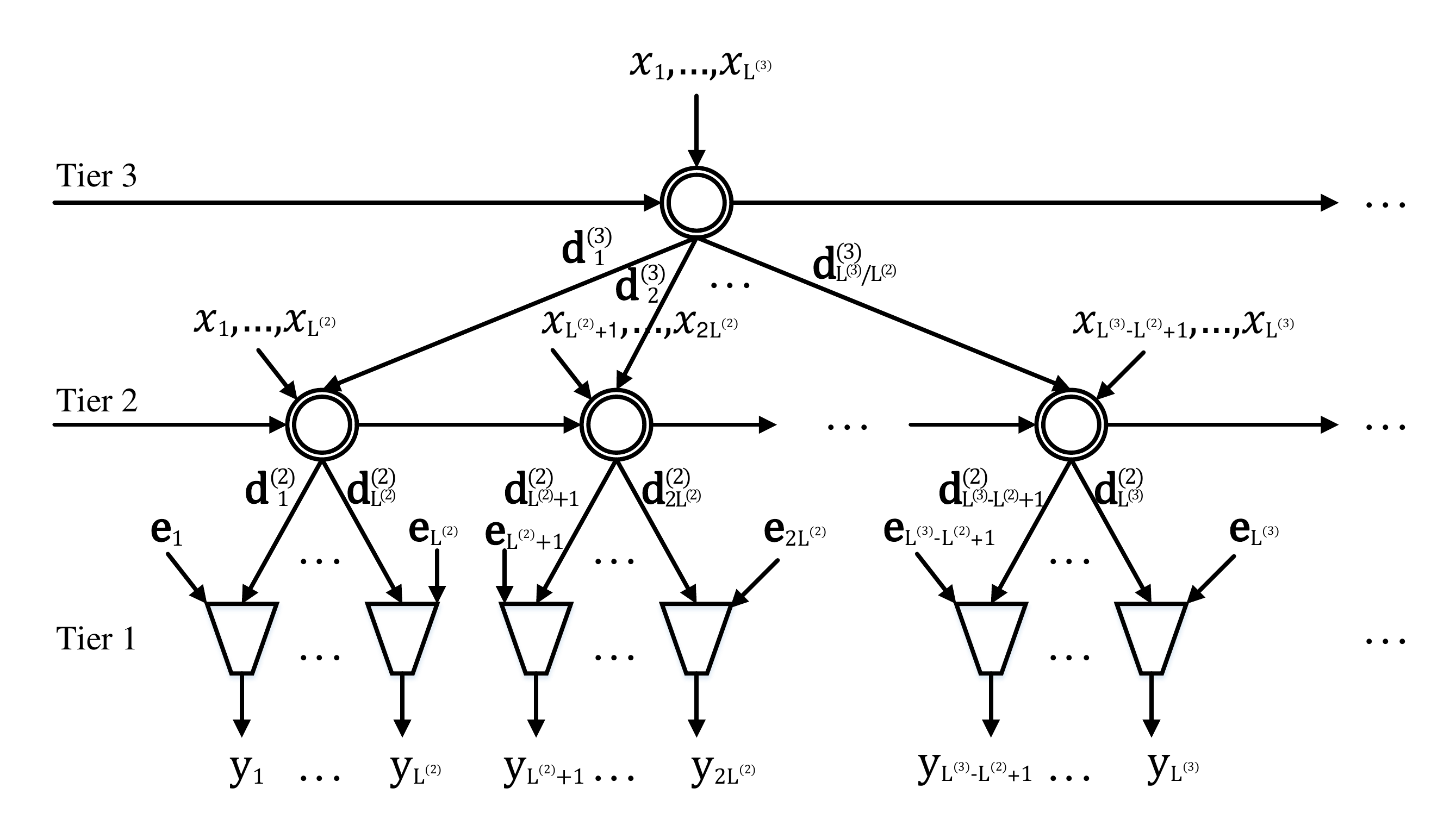}
    \caption{The structure of HRNNs for BWE, where concentric circles represent LSTM layers and inverted trapezoids represent FF layers.}
    \label{fig5: HRNNs}
\end{figure}

The structure of HRNNs for BWE is illustrated in Fig. \ref{fig5: HRNNs}.
Similar to SRNNs mentioned in Section \ref{subsec3A: Sample-Level Recurrent Neural Networks}, HRNNs are also composed of LSTM layers and FF layers.
Different from the plain LSTM-RNN structure of SRNNs, these LSTM and FF layers in HRNNs form a hierarchical structure of multiple tiers
and  each tier operates at a specific temporal resolution.
The bottom tier (i.e., Tier 1 in Fig. \ref{fig5: HRNNs}) deals with individual samples and outputs sample-level predictions.
Each higher tier operates on a lower temporal resolution (i.e., dealing with more samples per time step).
Each tier conditions on the tier above it except the top tier.
This model structure is similar to SampleRNN \cite{mehri2016samplernn}.
The main difference is that the original SampleRNN model is an unconditional audio generator which employs the history of output waveforms as network input and generates output waveforms in an autoregressive way.
While, the HRNN model shown in Fig. \ref{fig5: HRNNs} describes the mapping relationship between two waveform sequences directly without considering the autoregressive property of output waveforms.
This HRNN structure is specifically designed for BWE because narrowband waveforms are used as inputs in this task.
Removing autoregressive connections can help reduce the computation complexity and facilitate parallel computing at generation time.
Although conditional SampleRNNs have been developed and used as neural vocoders to reconstruct speech waveforms from acoustic parameters \cite{sotelo2017char2wav},
they still follow the autoregressive framework and are different from HRNNs.

Assume an HRNN has $K$ tiers in total (e.g., $K=3$ in Fig. \ref{fig5: HRNNs}).
Tier 1 works at sample-level and the other $K-1$ tiers are frame-level tiers since they operate at a temporal resolution lower than samples.

\subsubsection{Frame-level tiers}
\label{subsubsec3B1: Frame-level tiers}

The $k$-th tier $(1<k\leq K)$ operates on frames composed of $L^{(k)}$ samples. 
The range of time step at the $k$-th tier, $t^{(k)}$, is determined by  $L^{(k)}$.
Denoting the quantized input waveforms as $\bm{x}=[x_1,x_2,\dots,x_T]$ and assuming that $L$ represents the sequence length of $\bm{x}$ after zero-padding so that $L$ can be divisible by $L^{(K)}$, we can get
\begin{align}
\label{equ: Range of values}
t^{(k)}\in T^{(k)}=\{1,2,\dots,\frac{L}{L^{(k)}}\},1<k\leq K.
\end{align}
Furthermore, the relationship of temporal resolution between the $m$-th tier and the $n$-th tier ($1<m<n\leq K$) can be described as
\begin{align}
\label{equ: Relations between Tn and Tm}
T^{(n)}=\{t^{(n)}|t^{(n)}=\lceil\frac{t^{(m)}}{L^{(n)}/L^{(m)}}\rceil,t^{(m)}\in T^{(m)}\},
\end{align}
where $\lceil\cdot\rceil$ represents the operation of rounding up. 
It can be observed from (\ref{equ: Relations between Tn and Tm}) that one time step of the $n$-th tier corresponds to $L^{(n)}/L^{(m)}$ time steps of the $m$-th tier.
The frame inputs $\bm{{f}}_t^{(k)}$ at the $k$-th tier $(1<k\leq K)$ and the $t$-th time step can be written by framing and concatenation operations as
\begin{align}
\label{equ: Frame inputs in frame tiers}
\bm{\tilde{f}}_t^{(k)}&=[x_{(t-1)L^{(k)}+1},\dots,x_{tL^{(k)}}]^\top,\\
\bm{{f}}_t^{(k)}&=[\bm{\tilde{f}}_t^{(k)\top},...,\bm{\tilde{f}}_{t+c^{(k)}-1}^{(k)\top}]^\top,
\end{align}
where $t\in T^{(k)}$, $\bm{\tilde{f}}_t^{(k)}$ denotes the $t$-th waveform frame at the $k$-th tier, and
$c^{(k)}$ is the number of concatenated frames at the $k$-th tier.
We have  $c^{(3)}=c^{(2)}=1$ in the model structure shown in Fig. \ref{fig5: HRNNs}.

As shown in Fig. \ref{fig5: HRNNs}, the frame-level ties are composed of LSTM layers.
For the top tier (i.e., $k=K$), 
the LSTM units update their hidden states $\bm{h}_t^{(K)}$ based on the hidden states of previous time step $\bm{h}_{t-1}^{(K)}$ and the input at current time step $\bm{f}_t^{(K)}$.
If there is only one LSTM layer in the $K$-th tier, the calculation process can be formulated as
\begin{align}
\label{equ: Computation of RNNs in top tier}
\bm{h}_t^{(K)}={\cal H}(\bm{h}_{t-1}^{(K)},\bm{f}_t^{(K)}),t\in T^{(K)}.
\end{align}
If the top tier is composed of multiple LSTM-RNN layers, the hidden states can be calculated layer-by-layer iteratively.

Due to the different temporal resolution at different tiers,
the top tier generates $r^{(K)}=L^{(K)}/L^{(K-1)}$ conditioning vectors for the $(K-1)$-th tier at each time step $t\in T^{(K)}$.
This is implemented by producing a set of $r^{(K)}$ separate linear projections of $\bm{h}_t^{(K)}$ at each time step.
For the intermediate tiers (i.e., $1<k<K$), the processing of generating conditioning vectors is the same as that of the top tier.
Thus, we can describe the conditioning vectors uniformly as
\begin{align}
\label{equ: Conditioning vectors in frame tiers}
\bm{d}_{(t-1)r^{(k)}+j}^{(k)}=\bm{W}_j^{(k)}\bm{h}_t^{(k)}, j=1,2,\dots,r^{(k)},t\in T^{(k)},
\end{align}
where $1<k\leq K$ and $r^{(k)}=L^{(k)}/L^{(k-1)}$.

The input vectors of the LSTM layers at intermediate tiers are different from that of the top tier.
For the $k$-th tier ($1<k<K$), the input vector $\bm{i}_t^{(k)}$ at the $t$-th time step is composed by a linear combination of the frame inputs $\bm{f}_t^{(k)}$ and the conditioning vectors $\bm{d}_t^{(k+1)}$ given by the $(k+1)$-th tier as
\begin{align}
\label{equ: Inputs of RNNs at intermediate tiers}
\bm{i}_t^{(k)}=\bm{W}^{(k)}\bm{f}_t^{(k)}+\bm{d}_t^{(k+1)},t\in T^{(k)},
\end{align}
Thus, the output of the LSTM layer at the $k$-th tier ($1<k<K$) can be calculated as
\begin{align}
\label{equ: Compuation of RNNs at intermediate tiers}
\bm{h}_t^{(k)}={\cal H}(\bm{h}_{t-1}^{(k)},\bm{i}_t^{(k)}),t\in T^{(K)}.
\end{align}

\subsubsection{Sample-level tier}
\label{subsubsec3B2: Sample-level tier}

The sample-level tier (i.e., Tier 1 in Fig. \ref{fig5: HRNNs}) gives the probability distribution of the output sample $y_t$ conditioned on the current input sample $x_t$ (i.e., $L^{(1)}=1$) together with the conditioning vector $\bm{d}_t^{(2)}$ passed from the above tier which encodes history information of the input sequence, where $t\in T^{(1)}=\{1,2,\dots,\frac{L}{L^{(1)}}\}$. 
Since $x_t$ and $y_t$ are individual samples, it is convenient to model the correlation among them using a memoryless structure such as FF layers.
First, $x_t$ is mapped into a real-valued vector $\bm{e}_t$ by an embedding layer.
These embedding vectors form the input at each time step of the sample-level tier, i.e.,
\begin{align}
\label{equ: Frame inputs in sample tier}
\bm{f}_t^{(1)}=[\bm{e}_t^\top,...,\bm{e}_{t+c^{(1)}-1}^\top]^\top,
\end{align}
where $t\in T^{(1)}$,  $c^{(1)}$ is the number of concatenated sample embeddings at the sample-level tier.
In the model structure shown in Fig. \ref{fig5: HRNNs}, $c^{(1)}=1$.
Then, the input of the FF layers is a linear combination of $\bm{f}_t^{(1)}$ and $\bm{d}_t^{(2)}$ as
\begin{align}
\label{equ: Inputs of RNNs at tier 1}
\bm{i}_t^{(1)}=\bm{W}^{(1)}\bm{f}_t^{(1)}+\bm{d}_t^{(2)}, t\in T^{(1)}.
\end{align}

Finally, we can obtain the conditional probability distribution of the output sample $y_t$ by passing $\bm{i}_t^{(1)}$ through the FF layers.
The activation function of the last FF layer is a softmax function.
The output of FF layers describes the conditional distribution
\begin{align}
\label{equ: Conditional probability distribution of HRNNs}
p(y_t|x_1,x_2,\dots,x_{(\lceil \frac{t}{L^{(K)}}\rceil+c^{(K)}-1) L^{(K)}})=FF(\bm{i}_t^{(1)}),
\end{align}
where $t\in T^{(1)}$.

It is worth mentioning that the structure shown in Fig. \ref{fig5: HRNNs} is  non-casual which utilizes future input samples together with current and previous input samples
to predict current output sample (e.g., using $x_1,\dots,x_{L^{(3)}}$ to predict $y_1$ in Fig. \ref{fig5: HRNNs}).
Generally speaking, at most $c^{(K)}L^{(K)}-1$ input samples after the current time step are necessary in order to predict current output sample accroding to (\ref{equ: Conditional probability distribution of HRNNs}).
This is also a difference between our HRNN model and SampleRNN, which has a causal and autoregressive structure.

Similar to SRNNs, the parameters of HRNNs are estimated using cross-entropy cost function given a training set with parallel input and output sample sequences.
At generation time, each $y_t$ is predicted using the conditional probability distribution in (\ref{equ: Conditional probability distribution of HRNNs}).


\subsection{Conditional Hierarchical Recurrent Neural Networks}
\label{subsec3C: Conditional Hierarchical Recurrent Neural Networks}

\begin{figure}[t]
    \centering
    \includegraphics[height=6.5cm]{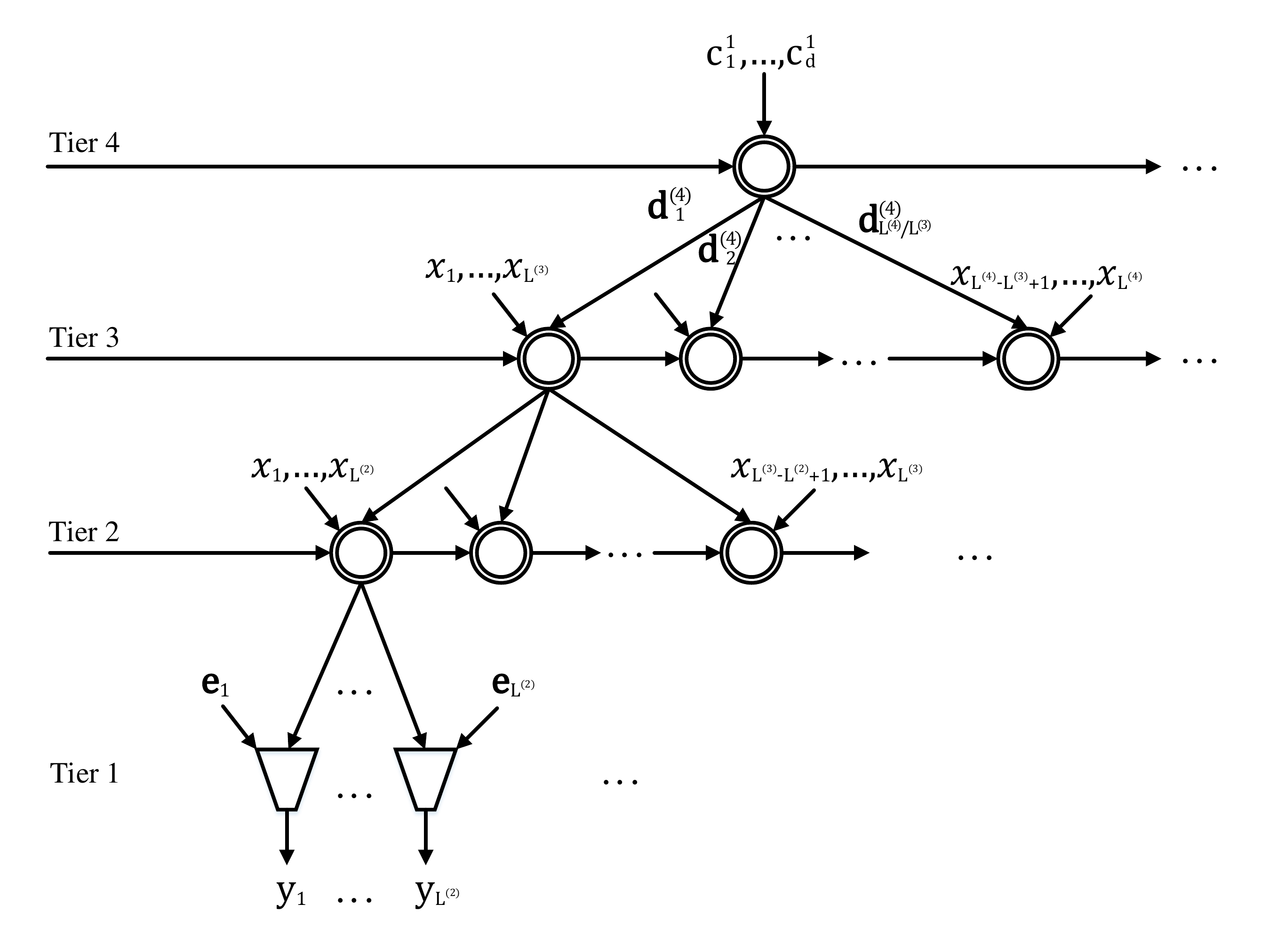}
    \caption{The structure of conditional HRNNs for BWE, where concentric circles represent LSTM layers and inverted trapezoids represent FF layers.}
    \label{fig6: Conditional HRNNs}
\end{figure}

Some frame-level auxiliary features extracted from input narrowband waveforms, such as bottleneck (BN) features \cite{yu2011improved},
have shown their effectiveness in improving the performance of vocoder-based BWE \cite{gu2016speech}.
In order to combine such auxiliary inputs with the HRNN model introduced in Section \ref{subsec3B: Hierarchical Recurrent Neural Networks},
a conditional HRNN structure is designed as shown in Fig. \ref{fig6: Conditional HRNNs}.


Compared with HRNNs, conditional HRNNs add an additional tier named \emph{conditional tier} on the top.
The input features of the conditional tier are frame-level auxiliary feature vectors extracted from input waveforms rather than waveform samples.
Assume the total number of tiers in a conditional HRNN is $K$ (e.g., $K=4$ in Fig. \ref{fig6: Conditional HRNNs})
and let $L^{(K)}$ donate the frame shift of auxiliary input features.
The equations (\ref{equ: Range of values}) and (\ref{equ: Relations between Tn and Tm}) in Section \ref{subsec3B: Hierarchical Recurrent Neural Networks} still works here.
Similar to the introductions in Section \ref{subsec3B: Hierarchical Recurrent Neural Networks}, the frame inputs at the conditional tier can be written as
\begin{align}
\label{equ: Frame inputs at top tier of conditional HRNNs}
\bm{c}_t=[c_1^t,c_2^t,\dots,c_d^t],t\in T^{(K)},
\end{align}
where $c_d^t$ represents the $d$-th dimension of the auxiliary feature vector at time $t$.
Then the calculations of (\ref{equ: Computation of RNNs in top tier})-(\ref{equ: Inputs of RNNs at tier 1}) for HRNNs are followed.
Finally, the conditional probability distribution for generating $y_t$ can be written as
\begin{align}
\label{equ: Conditional probability distribution of conditional HRNNs}
\nonumber p(y_t|x_1,\dots,x_{(\lceil \frac{t}{L^{(K)}}\rceil+c^{(K)}-1) L^{(K)}},&\bm{c}_1,\bm{c}_2,\dots,\bm{c}_{\lceil \frac{t}{L^{(K)}}\rceil})\\
&=FF(\bm{i}_t^{(1)}),
\end{align}
where $t\in T^{(1)}$, $\{\bm{c}_1,\bm{c}_2,\dots,\bm{c}_{\lceil \frac{t}{L^{(K)}}\rceil}\}$ are additional conditions introduced by the auxiliary input features.

\subsection{BWE Using SRNNs and HRNNs}
\label{subsec3D: BWE Using HRNNs}

\begin{figure}[t]
    \centering
    \includegraphics[height=5.5cm]{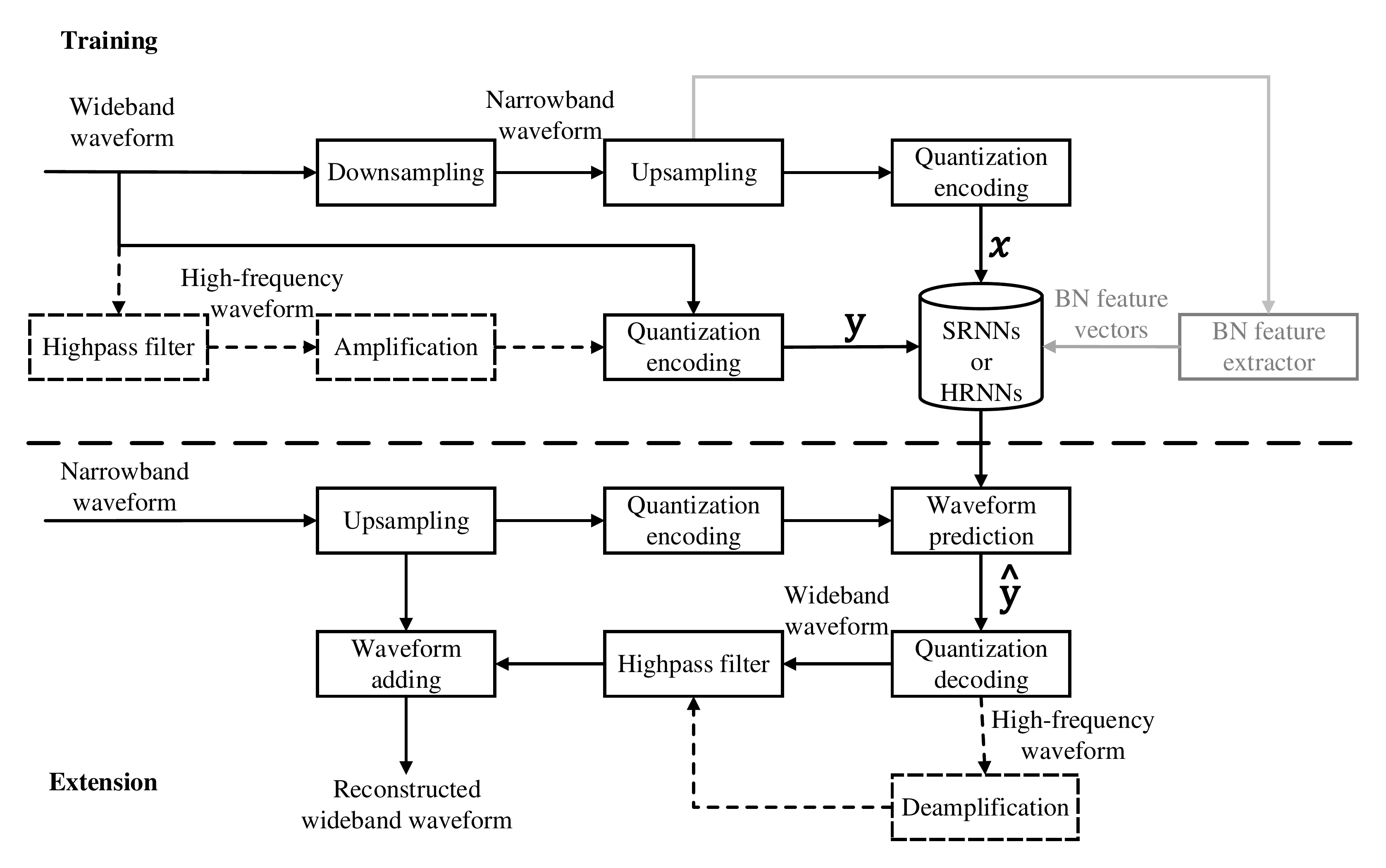}
    \caption{The flowchart of our proposed BWE methods.}
    \label{fig7: Flowchart of BWE}
\end{figure}

The flowchart of BWE using SRNNs or HRNNs are illustrated in Fig. \ref{fig7: Flowchart of BWE}.
There are two mapping strategies. One is to map the narrowband waveforms towards their corresponding wideband counterparts (named \emph{WB} strategy in the rest of this paper)
and the other is to map the narrowband waveforms towards the waveforms of the  high-frequency component of wideband speech (named \emph{HF} strategy).

A database with wideband speech recordings is used for model training.
At the training stage, the input narrowband waveforms are obtained by downsampling the 
wideband waveforms.
To guarantee the length consistency between the input and output sequences, the narrowband waveforms are then upsampled to the sampling rate of the wideband speech with zero high-frequency components. 
The upsampled narrowband waveforms are used as the model input.
The output waveforms are either the 
unfiltered wideband waveforms (\emph{WB} strategy)
or the high-frequency waveforms (\emph{HF} strategy).
The high-frequency waveforms are obtained by sending wideband speech into a high-pass filter and an amplifier for reducing quantization noise as the dotted lines in Fig. \ref{fig7: Flowchart of BWE}.
Before the waveforms are used for model training, all the input and output waveform samples are discretized by 8-bit $\mu$-law quantization.
The model parameters of SRNNs or HRNNs are trained under cross-entropy (CE) criterion
which optimizes the classification accuracy of discrete output samples on training set.

At the extension stage, the upsampled and quantized narrowband waveforms are fed into the trained SRNNs or HRNNs to generate the probability distributions of output samples.
Then each output sample is obtained by selecting the quantization level with maximum posterior probability.
Later, the quantized output  samples are decoded into continuous values using the inverse mapping of $\mu$-law quantization.
A deamplification process is conducted for the HF strategy in order to compensate the effect of amplification at training time.
Finally, the generated waveforms are high-pass filtered and added with the input narrowband waveforms to generate the final wideband waveforms.

Particularly for conditional HRNNs, BN features are used as auxiliary input in our implementation  as shown by the gray lines in Fig. \ref{fig7: Flowchart of BWE}.
BN features can be regarded as a compact representation of both linguistic and acoustic information \cite{yu2011improved}.
Here, BN features are extracted by a  DNN-based state classifier, which has a bottleneck layer with smaller number of hidden units than that of other hidden layers.
The inputs of the DNN are mel-frequency cepstral coefficients (MFCC) extracted from narrowband speech and the outputs are the posterior probability of HMM states.
The DNN is trained under cross-entropy (CE) criterion and is used as the BN feature extractor at extension time. 


\section{Experiments}
\label{sec4: Experiments}
\subsection{Experimental Setup}
\label{subsec4A: Experimental Setups}

The TIMIT corpus \cite{garofolo1993darpa} which contained English speech from multi-speakers with 16kHz sampling rate and 16bits resolution was adopted in our experiments.
We chose 3696 and 1153 utterances to construct  the training set and validation set respectively.
Another 192 utterances from the speakers not included in the training set and validation set were used as the test set to evaluate the performance of different BWE methods.
In our experiments, the narrowband speech waveforms sampled at 8kHz were obtained by downsampling the 
 wideband speech  at 16kHz.

Five BWE systems\footnote{Examples of reconstructed speech waveforms in our experiments can be found at \url{http://home.ustc.edu.cn/~ay8067/IEEEtran/demo.html}.}
were constructed for comparison in our experiments. The descriptions of these systems are as follows.

\begin{itemize}
\item \emph{\textbf{VRNN}}: Vocoder-based BWE method using LSTM-RNNs as introduced in Section \ref{subsec2B: Vocoder-based BWE Using LSTM-Based RNNs}.
    The \emph{DRNN-BN} system in \cite{gu2016speech} was used here for comparison, which predicted the LMS of high-frequency components using a deep LSTM-RNN with auxiliary BN features.
    Backpropagation through time (BPTT) algorithm was used to train the LSTM-RNN model based on the minimum mean square error (MMSE) criterion.
    In this system, a DNN-based state classifier was built to extract BN features.
    11-frames of 39-dimensional narrowband MFCCs 
    were used as the input of the DNN classifier and the posterior probabilities of 183 HMM states for 61 monophones were regarded as the output of the DNN classifier.
    The DNN classifier adopt 6 hidden layers where there were 100 hidden units at the BN layer and 1024 hidden units at other hidden layers. The BN layer was set as the fifth hidden layer so that the extractor could capture more linguistic information. This BN feature extractor was also used in the \emph{\textbf{CHRNN}} system.

\item \emph{\textbf{DCNN}}: Waveform-based BWE method using stacked dilated CNNs as introduced in Section \ref{subsec2C: Waveform-Based BWE Using Stacked Dilated CNNs}. 
    The \emph{CNN2-HF} system in \cite{gu2017speech} was used here for comparison, which predicted high-frequency waveforms using non-causal CNNs and performed better than other configurations.
\item \emph{\textbf{SRNN}}: Waveform-based BWE method using sample-level RNNs as introduced in Section \ref{subsec3A: Sample-Level Recurrent Neural Networks}. 
    The built model had two LSTM layers and two FF layers.
    Both the LSTM layers and the FF layers had 1024 hidden units and the embedding size was 256.
    The model was trained by stochastic gradient decent with a mini-batch size of 64 to minimize the cross entropy between the predicted and real probability distribution.
    Zero-padding was applied to make all the sequences in a mini-batch have the same length and the cost values of the added zero samples were ignored when computing the gradients.
    An \emph{Adam} optimizer \cite{kingma2014adam} was used to update the parameters with an initial learning rate 0.001.
    Truncated backpropagation through time (TBPTT) algorithm 
    was  employed to improve the efficiency of model training 
    and the truncated length was set to 480.
\item \emph{\textbf{HRNN}}: Waveform-based BWE method using HRNNs as introduced in Section \ref{subsec3B: Hierarchical Recurrent Neural Networks}. 
    The HRNN was composed of 3 tiers with two FF layers in Tier 1 and one LSTM layer each in Tier 2 and 3.
    Therefore, there were two LSTM layers and two FF layers in total which was the same as the \emph{\textbf{SRNN}} system.
    The number of $c^{(k)},k=\{1,2,3\}$ in (14) and (19) were set as $c^{(3)}=c^{(2)}=2,c^{(1)}=L^{(2)}$ in our experiments after tuning on the validation set.
    Some other setups, such as the dimension of the hidden units and the training method, were the same as that of the \emph{\textbf{SRNN}} system mentioned above.
    The frame size configurations of the HRNN model will be discussed in Section \ref{subsec4B: Structure and mapping type comparison for HRNNs}.
\item \emph{\textbf{CHRNN}}:  Waveform-based BWE method using conditional HRNNs as introduced in Section \ref{subsec3C: Conditional Hierarchical Recurrent Neural Networks}. 
    The BN features extracted by the DNN state classifier used by the \emph{\textbf{VRNN}} system were adopted as auxiliary conditions.
    The model was composed of 4 tiers.
    The top conditional tier had one LSTM layer with 1024 hidden units and the other three tiers were the same as the \emph{\textbf{HRNN}} system.
    Some basic setups and the training method were the same as the \emph{\textbf{HRNN}} system.
    The setup of the conditional tier will be introduced in detail in Section \ref{subsec4D: Exploration of the Effect of Additional Conditions on HRNNs}.
\end{itemize}


In our experiments, we first investigated the influence of frame sizes and mapping strategies (i.e., the \emph{WB} and \emph{HF} strategies introduced in Section \ref{subsec3D: BWE Using HRNNs}) on the performance of the \emph{\textbf{HRNN}} system.
Then, the comparison between different waveform-based BWE methods including the \emph{\textbf{DCNN}}, \emph{\textbf{SRNN}} and \emph{\textbf{HRNN}} systems was carried out.
Later, the effect of introducing BN features to HRNNs was studied by comparing the \emph{\textbf{HRNN}} system and the \emph{\textbf{CHRNN}} system.
Finally, our proposed waveform-based BWE method was compared with the conventional vocoder-based one. 

\subsection{Effects of Frame Sizes on HRNN-Based BWE}
\label{subsec4B: Structure and mapping type comparison for HRNNs}

\begin{figure}[t]
    \centering
    \includegraphics[height=6.5cm]{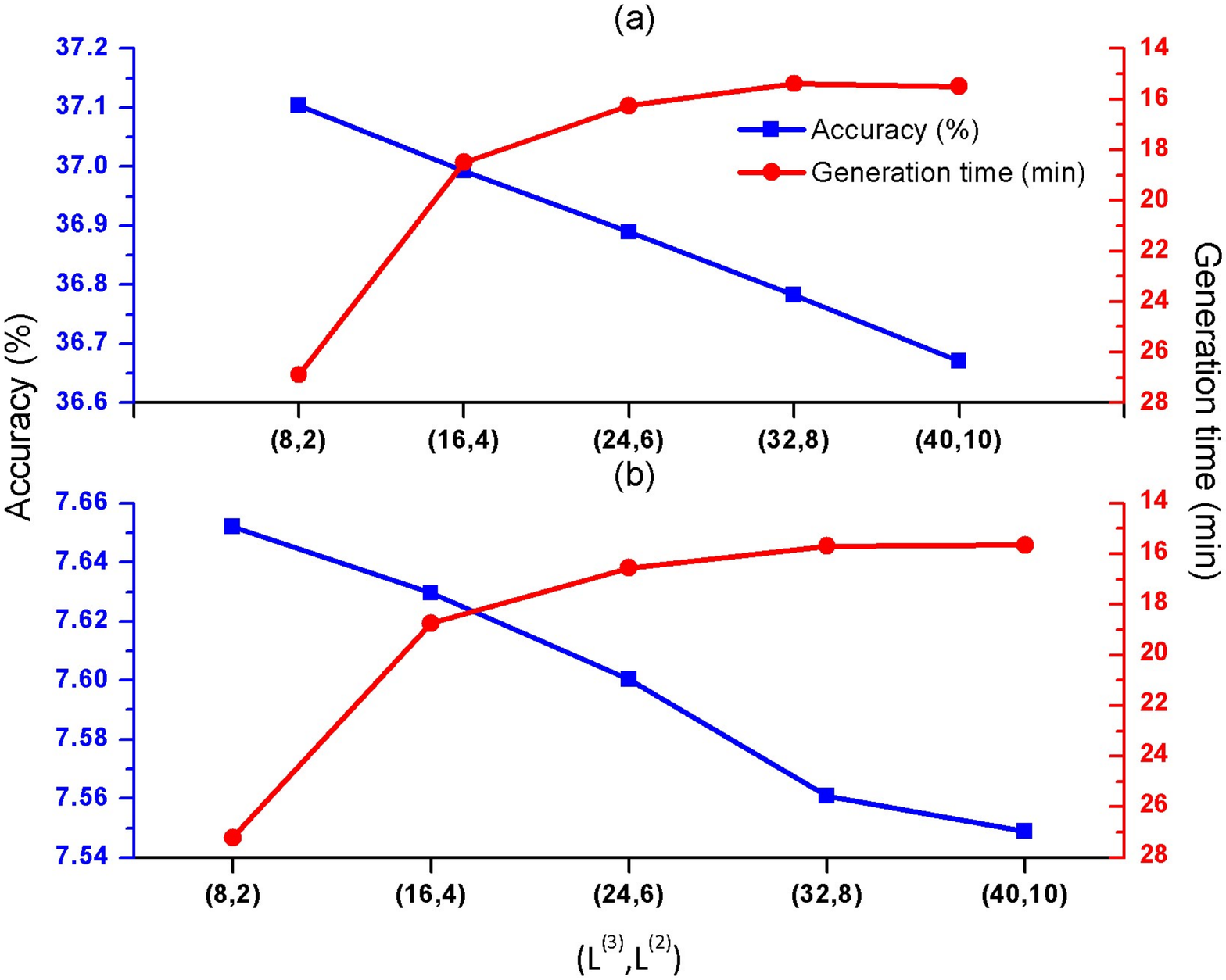}
    \caption{Accuracy and efficiency comparison for HRNN-based BWE with different $(L^{(3)},L^{(2)})$ configurations and using (a) \emph{WB} and (b) \emph{HF} mapping strategies.
    }
    \label{fig10: Accuracy and efficiency comparison for different L3 L2 combination of HRNNs}
\end{figure}

As introduced in Section \ref{subsec3B: Hierarchical Recurrent Neural Networks}, the frame sizes $L^{(k)}$ are key parameters that makes a HRNN model different from the conventional sample-level RNN.
In this experiment, we studied the effect of $L^{(k)}$ on the performance of HRNN-based BWE.
The HRNN models with several configurations of $(L^{(3)},L^{(2)})$ were trained and their accuracy and efficiency were compared as shown in  Fig. \ref{fig10: Accuracy and efficiency comparison for different L3 L2 combination of HRNNs}.
Here, the classification accuracy of predicting discrete waveform samples in the validation set was used to measure the accuracy of different models.
The total time of generating 1153 utterances in the validation set with mini-batch size of 64 on a single Tesla K40 GPU was used to measure the run-time efficiency.
Both the \emph{WB} and \emph{HF} mapping strategies were considered in this experiment.
From the results shown in Fig. \ref{fig10: Accuracy and efficiency comparison for different L3 L2 combination of HRNNs},
we can see that there existed conflict between the accuracy and the efficiency of the trained HRNN models.
Using smaller frame sizes of $(L^{(3)},L^{(2)})$ improved the accuracy of sample prediction while increased the computational complexity at the extension stage for both the \emph{WB} and \emph{HF}  strategies.
Finally, we chose $(L^{(3)},L^{(2)})=(16,4)$ as a trade-off and used this configuration for building the \emph{\textbf{HRNN}} system in the following experiments.

\subsection{Effects of Mapping Strategy on HRNN-Based BWE}
\label{subsec4B: Effects of Mapping Strategy on HRNN-based BWE}

It can be observed from Fig. \ref{fig10: Accuracy and efficiency comparison for different L3 L2 combination of HRNNs}
that the \emph{HF} strategy achieved much lower classification accuracy than the \emph{WB} strategy.
It is reasonable since it is more difficult to predict the aperiodic and noise-like high-frequency waveforms than to predict wideband waveforms.
Objective and subjective evaluations were conducted to investigate which strategy can achieve better performance for the HRNN-based BWE.


Since it is improper to compare the classification accuracy of these two strategies directly,
the score of Perceptual Evaluation of Speech Quality (PESQ) for wideband speech (ITU-T P.862.2) \cite{rec2007p}
was adopted as the objective measurement here.
We utilized the clean wideband speech as reference and calculated the PESQ scores of the 192 utterances in the test set generated using \emph{WB} and \emph{HF} strategies
(i.e., the \emph{\textbf{HRNN-WB}} system and the \emph{\textbf{HRNN-HF}} system) respectively.
For comparison, the PESQ scores of the upsampled narrowband utterances (i.e., with empty high-frequency components) were also calculated.
The average PESQ scores and their 95\% confidence intervals are shown in Table \ref{tab2: PESQ scores for HRNN-WB and HRNN-HF}.
The differences between any two of the three systems were significant according to the results of paired $t$-tests ($p<0.001$).
From Table \ref{tab2: PESQ scores for HRNN-WB and HRNN-HF}, we can see that the \emph{HF} strategy achieved higher PESQ score than the \emph{WB} strategy.
The average PESQ of the \emph{\textbf{HRNN-WB}} system was even lower than that of the upsampled narrowband speech.
This may be attributed to that the model in the \emph{\textbf{HRNN-WB}} system aimed to reconstruct the whole wideband waveforms and was incapable of generating high-frequency components as accurately as the \emph{\textbf{HRNN-HF}} system.

\begin{table}
\centering
    \caption{Average PESQ scores with 95\% confidence intervals on the test set when using \emph{WB} and \emph{HF} mapping strategies for HRNN-based BWE.}
    \begin{tabular}{c c c c}
        \hline
        \hline
         & \emph{\textbf{Narrowband}} & \emph{\textbf{HRNN-WB}}& \emph{\textbf{HRNN-HF}} \\
         \hline
         PESQ score &  3.63$\pm$0.0636 & 3.53$\pm$ 0.0438 &\textbf{3.75$\pm$ 0.0456}\\
        \hline
        \hline
    \end{tabular}
\label{tab2: PESQ scores for HRNN-WB and HRNN-HF}
\end{table}

\begin{figure}[t]
    \centering
    \includegraphics[height=6.5cm]{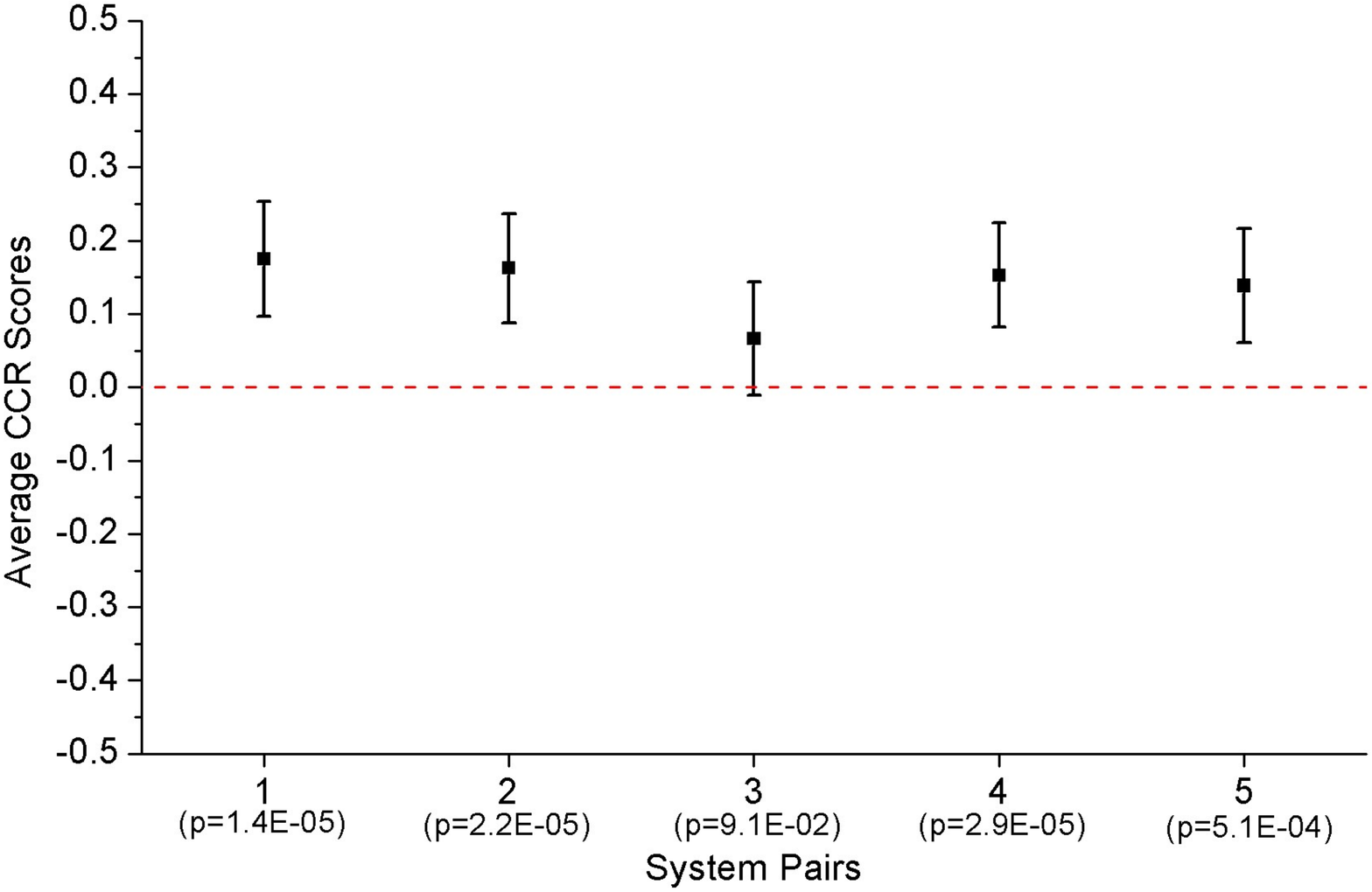}
    \caption{Average CCR scores of comparing five system pairs, including (1) \emph{\textbf{HRNN-HF}} vs. \emph{\textbf{HRNN-WB}}, (2)  \emph{\textbf{HRNN}} vs. \emph{\textbf{DCNN}}, (3) \emph{\textbf{HRNN}} vs. \emph{\textbf{SRNN}}, (4) \emph{\textbf{CHRNN}} vs. \emph{\textbf{HRNN}}, and (5) \emph{\textbf{CHRNN}} vs. \emph{\textbf{VRNN}}. The error bars represent 95\% confidence intervals and the numerical values in parentheses represent the $p$-value of one-sample $t$-test for different system pairs.}
    \label{fig: CCR}
\end{figure}

A 3-point comparison category rating (CCR) \cite{watson2001assessing} test was conducted on the Amazon Mechanical Turk (AMT) crowdsourcing platform (\url{https://www.mturk.com}) to compare the subjective performance of the \emph{\textbf{HRNN-WB}} and \emph{\textbf{HRNN-HF}} systems.
The wideband waveforms of 20 utterances randomly selected from the test set were reconstructed by the \emph{\textbf{HRNN-WB}} and \emph{\textbf{HRNN-HF}} systems.
Each pair of generated wideband speech were evaluated in random order by  15 native English listeners after rejecting improper listeners based on anti-cheating considerations \cite{buchholz2011crowdsourcing}.
The listeners were asked to judge which utterance in each pair had better speech quality or there was no preference.
Here, the \emph{\textbf{HRNN-WB}} system was used as the reference system.
The CCR scores of +1, -1, and 0 denoted that the wideband utterance reconstructed by the evaluated system, i.e., the \emph{\textbf{HRNN-HF}} system, sounded
better than, worse than, or equal to the sample generated by the reference system in each pair.
We calculated the average CCR score and its 95\% confidence interval through all pairs of utterances listened by all listeners.
Besides, one-sample $t$-test was also conducted to judge whether there was a significant difference between the average CCR score and 0 (i.e., to judge whether there was a significant difference between two systems) by examining the $p$-value.
The results are shown as the first system pair in Fig. \ref{fig: CCR},
which suggests that the \emph{\textbf{HRNN-HF}} system outperformed the \emph{\textbf{HRNN-WB}} system significantly.
This is consistent with the results of comparing these two strategies when dilated CNNs were used to model waveforms for the BWE task \cite{gu2017speech}.
Therefore, the \emph{HF} strategy was adopted in the following experiments for building waveform-based BWE systems.

\subsection{Model Comparison for Waveform-Based BWE }
\label{subsec4C: Model comparison for unconditional waveform-based BWE methods}

The performance of three waveform-based BWE systems, i.e., the \emph{\textbf{DCNN}}, \emph{\textbf{SRNN}} and \emph{\textbf{HRNN}} systems, were compared by objective and subjective evaluations.
The accuracy and efficiency metrics used in Section \ref{subsec4B: Structure and mapping type comparison for HRNNs} and the PESQ score used in Section \ref{subsec4B: Effects of Mapping Strategy on HRNN-based BWE}
were adopted as objective measurements.
Besides, two extra metrics were adopted here, including signal-to-noise ratio (SNR) \cite{tamamori2017speaker} which measured the distortion of waveforms and  log spectral distance (LSD) \cite{tamamori2017speaker} which reflected the distortion in frequency domain.
The SNR and LSD for voiced frames (denoted by SNR-V and LSD-V) and unvoiced frames (denoted by SNR-U and LSD-U) were also calculated separately for each system.
For the fairness of efficiency comparison, we set the mini-batch size as 1 for all the three systems when generating utterances in the test set.
The time of generating 1 second speech (i.e., 16000 samples for 16kHz speech) using a Tesla K40 GPU was recorded as the measurement of efficiency in this experiment.


Table \ref{tab4: Accuracy, PESQ scores and generation time of DCNN, SRNN and HRNN} shows the objective performance of the three systems on the test set.
The 95\% confidence intervals were also calculated for all metrics except the generation time.
The results of paired $t$-tests indicated that the differences between any two of the three systems on all metrics were significant ($p<0.01$).
For accuracy and PESQ score, the \emph{\textbf{DCNN}} system was not as good as the other two systems.
The \emph{\textbf{HRNN}} system achieved the best performance on both accuracy and PESQ score.
For SNR, the \emph{\textbf{HRNN}} system and the \emph{\textbf{DCNN}} system achieved the best performance on voiced segments and unvoiced segments respectively.
For LSD, the \emph{\textbf{HRNN}} system achieved the lowest overall LSD and the lowest LSD of unvoiced segments.
On the other hand, the \emph{\textbf{DCNN}} system achieved the lowest LSD of voiced frames among the three systems. 
Considering that LSDs were calculated using only amplitude spectra while SNRs were influenced by both amplitude and phase spectra of the reconstructed waveforms, 
it can be inferred that the \emph{\textbf{HRNN}} system was better at restoring the phase spectra of voiced frames than the \emph{\textbf{DCNN}} system 
according to the SNR-V and LSD-V results of these two systems shown in Table \ref{tab4: Accuracy, PESQ scores and generation time of DCNN, SRNN and HRNN}. 
In terms of the efficiency, the generation time of the \emph{\textbf{SRNN}} system was more than 5 times longer than that of the \emph{\textbf{HRNN}} system
due to the sample-by-sample calculation at all layers in the SRNN structure as discussed in Section \ref{subsec3A: Sample-Level Recurrent Neural Networks}.
Also, the efficiency of the \emph{\textbf{DCNN}} system was slightly worse than that of the \emph{\textbf{HRNN}} system.
The results reveal that HRNNs can help improve both the accuracy and efficiency of SRNNs by modeling long-span dependencies among sequences using a hierarchical structure.

\begin{table}
\centering
    \caption{Objective performance of the \emph{\textbf{DCNN}}, \emph{\textbf{SRNN}} and \emph{\textbf{HRNN}} systems on the test set.}
    \begin{tabular}{c c c c}
        \hline
        \hline
         & \emph{\textbf{DCNN}}& \emph{\textbf{SRNN}} & \emph{\textbf{HRNN}}\\
         \hline
         Accuracy (\%) &7.18$\pm$0.336 & 7.40$\pm$0.387 &\textbf{7.52$\pm$0.388}\\
         PESQ score &3.62$\pm$0.0532 & 3.70$\pm$0.0477 & \textbf{3.75$\pm$0.0456}\\
         SNR (dB) &\textbf{19.06$\pm$0.5983} &18.95$\pm$0.6053 &19.00$\pm$0.6099\\
         SNR-V (dB) &26.14$\pm$0.7557&26.06$\pm$0.7648&\textbf{26.21$\pm$0.7716}\\
         SNR-U (dB) &\textbf{10.49$\pm$0.4094}&10.32$\pm$0.4126&10.26$\pm$0.4124\\
         LSD (dB) &8.46$\pm$0.122&8.61$\pm$0.136&\textbf{8.30$\pm$0.127}\\
         LSD-V (dB) &\textbf{7.71$\pm$0.172}&8.09$\pm$0.203&8.02$\pm$0.194\\
         LSD-U (dB) &9.34$\pm$0.124&9.19$\pm$0.124&\textbf{8.57$\pm$0.107}\\
         Generation time (s) & 3.97 & 19.39 & \textbf{3.61}\\
         \hline
        \hline
    \end{tabular}
\label{tab4: Accuracy, PESQ scores and generation time of DCNN, SRNN and HRNN}
\end{table}

\begin{figure}[t]
    \centering
    \includegraphics[height=9.5cm]{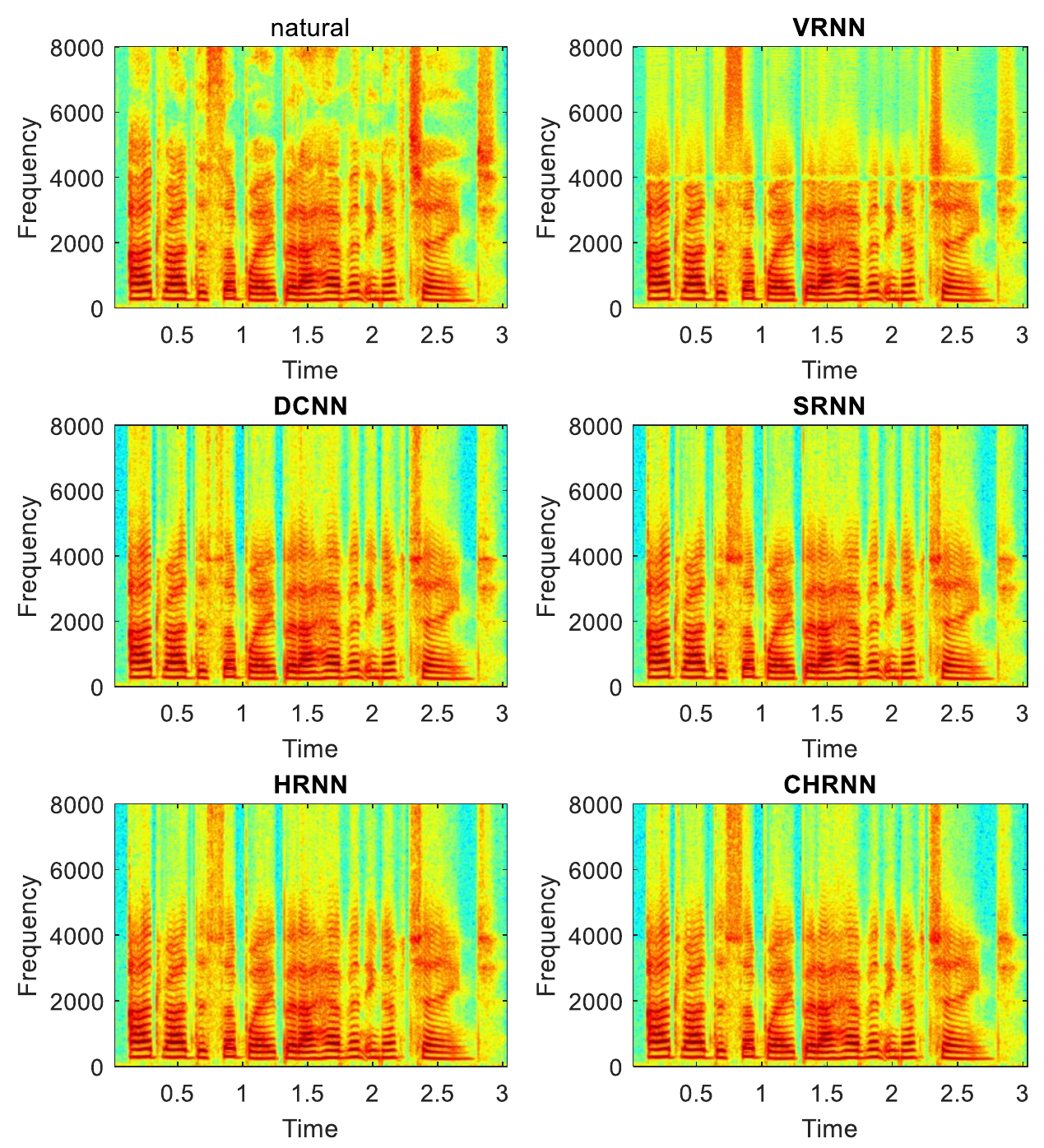}
    \caption{The spectrograms of clean wideband speech and the output of BWE using  five systems for an example sentence in the test set.}
    \label{fig12: Spectrogram of five systems}
\end{figure}

The spectrograms extracted from clean wideband speech and the output of BWE using  the \emph{\textbf{DCNN}}, \emph{\textbf{SRNN}} and \emph{\textbf{HRNN}} systems for an example sentence in the test set are shown in Fig. \ref{fig12: Spectrogram of five systems}.
It can be observed that the high-frequency energy of some unvoiced segments generated by the \emph{\textbf{DCNN}} system was much weaker than that of the natural speech and the outputs of the \emph{\textbf{SRNN}} and \emph{\textbf{HRNN}} systems.
Compared with the \emph{\textbf{SRNN}} and \emph{\textbf{HRNN}} systems, the \emph{\textbf{DCNN}} system was better at reconstructing the high-frequency harmonic structures of some voiced segments.
These observations are in line with the LSD results discussed earlier.

Furthermore, two 3-point CCR tests were carried out to evaluate the subjective performance of the \emph{\textbf{HRNN}} system by using the \emph{\textbf{DCNN}} system and the \emph{\textbf{SRNN}} system as the reference system respectively.
The configurations of the tests were the same as the ones introduced in Section \ref{subsec4B: Effects of Mapping Strategy on HRNN-based BWE}.
The results are shown as the second and third system pairs in Fig. \ref{fig: CCR}.
We can see that our proposed HRNN-based method generated speech with significantly better quality than the dilated CNN-based method.
Compared with the \emph{\textbf{SRNN}} system, the \emph{\textbf{HRNN}} system was slightly better while the superiority was insignificant at 0.05 significance level.
However, the \emph{\textbf{HRNN}} system was much more efficient than the \emph{\textbf{SRNN}} system at generation time as shown in Table \ref{tab4: Accuracy, PESQ scores and generation time of DCNN, SRNN and HRNN}.

\subsection{Effects of Additional Conditions on HRNN-Based BWE}
\label{subsec4D: Exploration of the Effect of Additional Conditions on HRNNs}

We compared the \emph{\textbf{HRNN}} system with the \emph{\textbf{CHRNN}} system by objective and subjective evaluations
to explore the effects of additional conditions on HRNN-based BWE.
As introduced in Section \ref{subsec4A: Experimental Setups}, the BN features were used as additional conditions in the \emph{\textbf{CHRNN}} system
since they can provide linguistic-related information besides the acoustic waveforms.
The \emph{\textbf{CHRNN}} system adopted the conditional HRNN structure introduced in Section \ref{subsec3C: Conditional Hierarchical Recurrent Neural Networks} with 4 tiers. The dimension of BN features was 100 and the frame size at the top conditional tier was $L^{(4)}=160$ because the frame shift of BN features was 10ms, corresponding to 160 samples for 16kHz speech.

The objective measurements used in Section \ref{subsec4C: Model comparison for unconditional waveform-based BWE methods} were adopted here
to compare the \emph{\textbf{HRNN}} and \emph{\textbf{CHRNN}} systems.
The results are shown in Table \ref{tab6: Accuracy, PESQ scores and generation time of HRNN and CHRNN}.
The \emph{\textbf{CHRNN}} system outperformed the \emph{\textbf{HRNN}} system on PESQ score while its prediction accuracy was not as good as the \emph{\textbf{HRNN}} system.
For SNR, these two systems achieved similar performance.
The results of LSD show that the \emph{\textbf{CHRNN}} system was better at reconstructing voiced frames and the \emph{\textbf{HRNN}} system was on the contrary.
In terms of efficiency, the generation time of the \emph{\textbf{CHRNN}} system was higher than that of the \emph{\textbf{HRNN}} system due to the extra conditional tier.

\begin{table}
\centering
    \caption{Objective performance of the \emph{\textbf{HRNN}} and \emph{\textbf{CHRNN}} systems on the test set together with the $p$ values of paired $t$-tests.}
    \begin{tabular}{c c c c}
        \hline
        \hline
         & \emph{\textbf{HRNN}}& \emph{\textbf{CHRNN}}&$p$-value\\
         \hline
         Accuracy (\%)&\textbf{7.52$\pm$0.388} & 7.46$\pm$0.385&$<$0.001\\
         PESQ score &3.75$\pm$0.0456 & \textbf{3.79$\pm$0.0394}&$<$0.001\\
         SNR (dB) &19.00$\pm$0.6099&18.99$\pm$0.5946&0.322\\
         SNR-V (dB) &\textbf{26.21$\pm$0.7716}&26.13$\pm$0.7539&$<$0.001\\
         SNR-U (dB) &10.26$\pm$0.4124&\textbf{10.34$\pm$0.4097}&$<$0.001\\
         LSD (dB) &8.30$\pm$0.127&8.27$\pm$0.123&0.301\\
         LSD-V (dB) &8.02$\pm$0.194&\textbf{7.89$\pm$0.185}&$<$0.001\\
         LSD-U (dB) &\textbf{8.57$\pm$0.107}&8.66$\pm$0.103&$<$0.01\\
         Generation time (s) & \textbf{3.61} & 4.17&--\\
         \hline
        \hline
    \end{tabular}
\label{tab6: Accuracy, PESQ scores and generation time of HRNN and CHRNN}
\end{table}

A 3-point CCR test was also conducted to evaluate the subjective performance of the \emph{\textbf{CHRNN}} system by using the  \emph{\textbf{HRNN}} system as the reference system
and following the evaluation configurations introduced in Section \ref{subsec4B: Effects of Mapping Strategy on HRNN-based BWE}.
The results are shown as the fourth system pairs in Fig. \ref{fig: CCR}, which reveal that
utilizing  BN features as additional conditions in HRNN-based BWE can improve the subjective quality of reconstructed wideband speech significantly.
Fig. \ref{fig12: Spectrogram of five systems} also shows the spectrogram of the wideband speech generated by the \emph{\textbf{CHRNN}} system for an example sentence.
Comparing the spectrograms produced by the \emph{\textbf{HRNN}} system and the \emph{\textbf{CHRNN}} system,
we can observe that the high-frequency components  generated by the \emph{\textbf{CHRNN}} system were stronger than the \emph{\textbf{HRNN}} system. This may lead to better speech quality as shown in Fig. \ref{fig: CCR}.


\subsection{Comparison between Waveform-Based and Vocoder-Based BWE Methods}
\label{subsec4E: Model Comparison for waveform-based and vocoder-based BWE methods}

Finally, we compared the performance of  vocoder-based and waveform-based BWE methods by
conducting objective and subjective evaluations between the \emph{\textbf{VRNN}} system and the \emph{\textbf{CHRNN}} system
since both systems adopted BN features as auxiliary input.
The objective results including PESQ, SNR and LSD are shown in Table \ref{tab7: objective evaluations of VRNN and CHRNN}.
The \emph{\textbf{CHRNN}} system achieved significantly better SNR than that of the \emph{\textbf{VRNN}} system,
which suggested that our proposed waveform-based method can restore the phase spectra more accurately than the conventional vocoder-based method.
For PESQ and LSD, the \emph{\textbf{CHRNN}} system was not as good as the \emph{\textbf{VRNN}} system.
This is reasonable considering that the \emph{\textbf{VRNN}} system modeled and predicted LMS directly which were used in the calculation of PESQ and LSD.
A 3-point CCR test was also conducted to evaluate the subjective performance of the \emph{\textbf{CHRNN}} system by using the  \emph{\textbf{VRNN}} system as the reference system
and following the evaluation configuratioins introduced in Section \ref{subsec4B: Effects of Mapping Strategy on HRNN-based BWE}.
The results are shown as the fifth system pairs in Fig. \ref{fig: CCR}.
We can see that the CCR score was high than 0 significantly which indicates that the \emph{\textbf{CHRNN}} system  can achieve significantly higher quality of
reconstructed wideband speech than the \emph{\textbf{VRNN}} system.

\begin{table}
\centering
    \caption{Objective performance of the \emph{\textbf{VRNN}} and \emph{\textbf{CHRNN}} systems on the test set together with the $p$ values of paired $t$-tests.}
    \begin{tabular}{c c c c}
        \hline
        \hline
         & \emph{\textbf{VRNN}}& \emph{\textbf{CHRNN}}&$p$ value\\
         \hline
         PESQ score &\textbf{3.87$\pm$0.0368} & 3.79$\pm$0.0394&$<$0.001\\
         SNR (dB) &17.76$\pm$0.6123&\textbf{18.99$\pm$0.5946}&$<$0.001\\
         SNR-V (dB) &25.00$\pm$0.7333&\textbf{26.13$\pm$0.7539}&$<$0.001\\
         SNR-U (dB) &9.01$\pm$0.424&\textbf{10.34$\pm$0.4097}&$<$0.001\\
         LSD (dB) &\textbf{6.69$\pm$0.110}&8.27$\pm$0.123&$<$0.001\\
         LSD-V (dB) &\textbf{6.86$\pm$0.148}&7.89$\pm$0.185&$<$0.001\\
         LSD-U (dB) &\textbf{6.45$\pm$0.0972}&8.66$\pm$0.103&$<$0.001\\
         \hline
        \hline
    \end{tabular}
\label{tab7: objective evaluations of VRNN and CHRNN}
\end{table}

Comparing the spectrograms produced by the \emph{\textbf{VRNN}} system and the \emph{\textbf{CHRNN}} system in Fig. \ref{fig12: Spectrogram of five systems},
it can be observed that the \emph{\textbf{CHRNN}} system performed better than the \emph{\textbf{VRNN}} system in generating the high-frequency harmonics for  voiced sounds.  
Besides, the high-frequency components generated by the \emph{\textbf{CHRNN}} system were less over-smoothed and more natural than that of the \emph{\textbf{VRNN}} system at unvoiced segments.
Furthermore,  there was a discontinuity between the low-frequency and high-frequency spectra of the speech generated the \emph{\textbf{VRNN}} system,
which was also found in other vocoder-based BWE method \cite{li2015deep}.
As shown in Fig. \ref{fig12: Spectrogram of five systems}, the waveform-based systems alleviated this discontinuity effectively.
These experimental results indicate the superiority of modeling and generating speech waveforms directly over utilizing vocoders for feature extraction and waveform reconstruction on the BWE task.

\subsection{Analysis and Discussion}
\label{subsec4F: Analysis and Discussion}
\emph{1) Maximal latency of different BWE systems}


\begin{table}
\centering
    \caption{Maximal latencies ($ms$) of the five BWE systems. The sampling rate of wideband waveforms is $f_s=16kHz$.}
    \resizebox{8.5cm}{1.7cm}{
    \begin{tabular}{c c c}
        \hline
        \hline
         & Maximal Latency & Remarks\\
         \hline
         \vspace*{1 mm}\emph{\textbf{VRNN}} &$WS=25$ & \tabincell{c}{$WS$: window size in $ms$ of STFT  \\for extracting spectral parameters.}\\
         \vspace*{1 mm}\emph{\textbf{DCNN}} &$\frac{N/2}{f_s}=32$ & $N+1$: length of receptive field.\\
         \vspace*{1 mm}\emph{\textbf{SRNN}} & $0$ & None\\
         \vspace*{1 mm}\emph{\textbf{HRNN}} & $\frac{c^{(3)}L^{(3)}-1}{f_s}=1.9375$ & \tabincell{c}{$c^{(3)}$, $L^{(3)}$: number of concatenated \\ frames, frame size at Tier 3.}\\
         \vspace*{1 mm}\emph{\textbf{CHRNN}} & $WS=25$ & \tabincell{c}{$WS$: window size in $ms$ of STFT  \\for extracting spectral parameters.}\\
         \hline
        \hline
    \end{tabular}}
\label{tab7: Maximal latency analysis of five BWE systems}
\end{table}

Some application scenarios have strict requirement on the latency of BWE algorithm.
We compared the maximal latency of the five BWE systems listed in Section \ref{subsec4A: Experimental Setups}
and the results are shown in  Table \ref{tab7: Maximal latency analysis of five BWE systems}.
Here, the latency refers to the duration of future input samples that are necessary for predicting current output sample.
The maximal latencies of the \emph{\textbf{VRNN}} system and the \emph{\textbf{CHRNN}} system were both determined by the window size of STFT for extracting LMS and MFCC parameters,
which was 25 ms in our implementation.
The maximal latencies of the other three systems depended on their structures.
The \emph{\textbf{SRNN}} system processed input waveforms and generate output waveforms sample-by-sample without latency according to (\ref{equ: Conditional probability of SRNNs}).
Because the non-causal CNN structure shown in Fig. \ref{fig2: DCNNs} was adopted by the \emph{\textbf{DCNN}} system
and its receptive field length was about 64\emph{ms} \cite{gu2017speech}, it made the highest latency among the five systems.
The latency of the \emph{\textbf{HRNN}} system was relatively short because the number of concatenated frames and the frame size of the top tier were small ($c^{(3)}=2$ and $L^{(3)}=16$).

\emph{2) Run-time efficiency of waveform-based BWE}

One deficiency of the waveform-based BWE methods is that they are very time-consuming at generation time.
As shown in Table \ref{tab4: Accuracy, PESQ scores and generation time of DCNN, SRNN and HRNN} and Table \ref{tab6: Accuracy, PESQ scores and generation time of HRNN and CHRNN},
the \emph{\textbf{HRNN}} system achieved the best run-time efficiency among the four waveform-based systems, which still took 3.61 seconds to generate 1 second speech in our current implementation.
Therefore, to accelerate the computation of HRNNs is an important task of our future work.
As shown in Fig. \ref{fig10: Accuracy and efficiency comparison for different L3 L2 combination of HRNNs},
using longer frame sizes may help reduce the computational complexity of HRNNs.
Another possible way is to reduce the number of hidden units and other model parameters similar to the attempt of accelerating WaveNet for speech synthesis \cite{arik2017deep}.
\section{Conclusion}
\label{sec5: Conclusion}

In this paper, we have proposed a novel waveform modeling and generation method using hierarchical recurrent neural networks (HRNNs) to fulfill the speech bandwidth extension (BWE) task.
HRNNs adopt a hierarchy of recurrent modules to capture long-span dependencies between input and output waveform sequences.
Compared with the plain sample-level RNN and the stacked dilated CNN, the proposed HRNN model achieves better accuracy and efficiency of predicting high-frequency waveform samples. 
Besides, additional conditions, such as the bottleneck features (BN) extracted from narrowband speech, can further improve subjective quality of reconstructed wideband speech. 
The experimental results show that our proposed HRNN-based method achieves higher subjective preference scores than the conventional vocoder-based method using LSTM-RNNs.
To evaluate the performance of our proposed methods using practical band-limited speech data, to improve the efficiency of waveform generation using HRNNs, and to utilize other types of additional conditions  will be the tasks of our future work.


\bibliographystyle{IEEEtran}
\bibliography{mybib}
\end{document}